\documentclass[10pt,conference]{IEEEtran}
\IEEEoverridecommandlockouts
\usepackage{enumerate}
\usepackage[T1]{fontenc}
\usepackage{xparse}
\usepackage{enumitem}
\usepackage{tabularx}
\usepackage{scalerel}
\usepackage{cite}
\usepackage{amsmath,amssymb,amsfonts}
\usepackage{algorithmic}
\usepackage{graphicx}
\usepackage{textcomp}
\usepackage{xcolor}
\usepackage{multirow}
\usepackage[colorinlistoftodos]{todonotes}
\usepackage{booktabs}
\usepackage{subcaption}
\usepackage[font=small,labelfont=bf,tableposition=top]{caption}
\usepackage[compact]{titlesec}
\usepackage{rotating}
\usepackage{array,makecell}

\titlespacing{\section}{0pt}{5pt}{0pt}
\titlespacing{\subsection}{0pt}{5pt}{0pt}
\titlespacing{\subsubsection}{0pt}{5pt}{0pt}
\AtBeginDocument{
  \setlength\abovedisplayskip{0pt}
  \setlength\belowdisplayskip{0pt}
}

\makeatletter
\newcolumntype{e}[1]{
   >{\minipage[t]{\linewidth}%
     \NoHyper
     \let\\\tabularnewline
     \enumerate
        \addtolength{\rightskip}{0pt plus 50pt}
        \setlength{\itemsep}{-\parsep}}%
   p{#1}%
   <{\@finalstrut\@arstrutbox\endenumerate
     \endNoHyper
     \endminipage}}

\newcolumntype{i}[1]{
   >{\minipage[t]{\linewidth}%
        \let\\\tabularnewline
        \itemize
           \addtolength{\rightskip}{0pt plus 50pt}%
           \setlength{\itemsep}{-\parsep}}%
   p{#1}%
   <{\@finalstrut\@arstrutbox\enditemize\endminipage}}

\AtBeginDocument{%
    \@ifpackageloaded{hyperref}{}%
        {\let\NoHyper\relax\let\endNoHyper\relax}}
\makeatother

\begin{document}
%

\title{Edit-Run Behavior in Programming and Debugging}

%

\author{\IEEEauthorblockN{Abdulaziz Alaboudi}
\IEEEauthorblockA{\textit{George Mason University} \\
Fairfax, Virginia, USA \\
aalaboud@gmu.edu}
\and
\IEEEauthorblockN{Thomas D. LaToza}
\IEEEauthorblockA{\textit{George Mason University} \\
Fairfax, Virginia, USA \\
tlatoza@gmu.edu}
}

\maketitle
\IEEEpubidadjcol

\begin{abstract}
As developers program and debug, they continuously edit and run their code, a behavior known as edit-run cycles. While techniques such as live programming are intended to support this behavior, little is known about the characteristics of edit-run cycles themselves. To bridge this gap, we analyzed 28 hours of programming and debugging work from 11 professional developers which encompassed 
over three thousand development activities. We mapped activities to edit or run steps, constructing 581 debugging and 207 programming edit-run cycles. We found that edit-run cycles are frequent. Developers edit and run the program, on average, 7 times before fixing a defect and twice before introducing a defect. Developers waited longer before again running the program when programming than debugging, with a mean cycle length of 3 minutes for programming and 1 minute for debugging. Most cycles involved an edit to a single file after which a developer ran the program to observe the impact on the final output. Edit-run cycles which included activities beyond edit and run, such as navigating between files, consulting resources, or interacting with other IDE features, were much longer, with a mean length of 5 minutes, rather than 1.5 minutes. We conclude with a discussion of design recommendations for tools to enable more fluidity in edit-run cycles.
\end{abstract}

%
%

%
%

%

\section{Introduction}
Development work can be characterized as a three step process in which developers edit, compile, and run code. In modern development environments, the compile step is automated, and this behavior is often referred to as an edit-run cycle. During the edit step, developers read and change the source code while adding new functionality or fixing a defect. During the run step, developers inspect and execute their program to test and observe the impact of the newly added change or patch. Developers' work often requires more than a single edit and single run step, requiring cycling through edit and run steps to achieve a desired output. Developers may repeat edit-run cycles during debugging to test different hypotheses\cite{BohmeFSE-DebuggingHypotheses, Zeller2005, Perscheid2017}. In programming, developers following test-driven development continuously introduce minor edits before again running their tests \cite{beck2003test}.

To support developers' work within edit-run cycles, researchers have designed a variety of tools. Live programming environments \cite{tanimoto2013perspective} aim to improve developers' productivity by merging the edit and run steps into a single step \cite{rein2018exploratory}, allowing developers to edit and run the program concurrently\cite{goldberg1983smalltalk,tanimoto1990viva, Tanimoto,mcdirmid2007living,lau2021tweakit, burckhardt2013it}. Live programming environments support developers within the edit step by generating code snippets using examples provided by developers \cite{ferdowsifard2020small, santolucito2019live} or by supporting direct manipulation of the output \cite{chugh2016programmatic, kery2020mage}. Rather than displaying only the final output of the program, live programming  systems support debugging by displaying intermediate values created during the program's execution~\cite{lieber2014addressing,oney2014interstate}.

Live programming environments envision improving the \textit{fluidity} of the programming experience by changing how developers work within edit-run cycles \cite{mcdirmid2016promise}. The end goal is to create fluid edit-run cycles which empower developers to focus on editing the program and immediately observing the impact of their edit in the output and intermediate state generated by the program. This fluid workflow envisioned by live programming environments has three core properties:

\begin{enumerate}
    \item Developers engage in short and frequent edit-run cycles.
    \item Developers focus on the edit step while observing the output.
    \item Developers repeat edit-run cycles sequentially without interruption within or between cycles. 
\end{enumerate}


Little is known about how close developers' current edit-run behavior is to this ideal. Embedded within their designs, live programming environments embody implicit assumptions about the behavior of developers within edit-run cycles. If developers were to work completely within a live programming environment, how would their workflow adapt?

To investigate these questions, we conducted the first empirical study of edit-run behavior. We used publicly available data to analyze 28 hours of work on open source projects by 11 professional developers  \cite{alaboudi2021exploratory}. Using this dataset, we constructed edit-run cycles that occurred in debugging and programming. We then investigated the fluidity developers are able to achieve within their edit-run cycles. Specifically, we examined:  

\begin{itemize}
\item[\textbf{RQ1}] How long and frequent are edit-run cycles?
 
\item[\textbf{RQ2}] How do developers edit and run during edit-run cycles?

\item[\textbf{RQ3}] How sequential are edit-run cycles, and what causes gaps within and between cycles?
 \end{itemize}

As developers' goals may influence their edit-run behavior, we analyzed and report findings separately for edit-run cycles in debugging and programming. Our analysis yielded 581 and 207 edit-run cycles in debugging and programming. 


We found that edit-run cycles in debugging lasted one minute on average and occurred seven times every 10 minutes. Cycles in programming were longer and less frequent, lasting on average three minutes and occurring twice every eight minutes. Developers who used a text editor completed 69\% more edit-run cycles in each debugging episode, which were each 55\% longer, than developers who used an IDE. Both debugging and programming edit-run cycles primarily focused on editing. Developers spent more than half of edit-run cycle time editing.  Developers edited only one file of code per cycle in 70\% of debugging and 60\% of programming cycles. These findings suggest that developers' edit-run cycle behavior largely achieves the first two proprieties of fluidity, with edit-run cycles in debugging closer to the ideal than programming.

Our analysis also revealed gaps which interrupted developers with other activities within and between edit-run cycles. Edit-run cycles with gaps were four times longer than edit-run cycles with no gaps. 
We identified four causes of these gaps, including working with scattered code, unfamiliar third party APIs, disintegrated development environments, and waiting to compile. We conclude with a set of design recommendations for programming tools which empower development work by improving fluidity.

\section {Background}


Our study of edit-run behavior builds on prior work examining debugging, programming, and exploratory programming as well as the ideal of a fluid live programming experience envisioned by live programming environments.



Edit-run cycles may occur while debugging, as developers formulate and test hypotheses about the cause of a defect \cite{Ko2007InformationNeeds,BohmeFSE-DebuggingHypotheses, Zeller2005, Perscheid2017}. To formulate a hypothesis, developers look for clues in the source code. To test a hypothesis, developers may edit the source code and run the program. Developers often test multiple incorrect hypotheses before fixing the defect~\cite{alaboudiHypotheses, BohmeFSE-DebuggingHypotheses, Ko2006TSE}, which may result in multiple edit-run cycles. Researchers have argued repeated edit-run cycles caused by testing fixes based on incorrect hypotheses may be avoided by helping developers debug more systematically \cite{zeller2001automated}. WhyLine, for example, systematically guides developers to navigate through the source code from the incorrect output to the source of the defect. In this way, WhyLine helps developers avoid formulating hypotheses around irrelevant parts of the code. Fault localization \cite{MarkWeiser1984ProgramSlicing, DeMillo1996, Zhang2006, XiangyuZhang2003} tools also allow developers to avoid formulating incorrect hypotheses by offering a ranked list of potential fault locations that developers need to investigate. In this way, developers might more quickly understand the cause of the defect and require fewer edit-run cycles to fix it.


While programming, developers may continuously run their programs while editing the source code to check their progress. One effective programming strategy is to divide the programming problem into sub-problems, allowing developers focus on sub-problems and test them individually \cite{Anton}. This approach has gained exposure and popularity in the form of test-driven development~\cite{beck2003test}. Developers test first, writing a unit test and only then write the minimal amount of code to make the test pass. After finishing the edit step, developers run the test case against the new code. This creates an edit-run cycle. Developers continue in edit-run cycles,  writing code and running unit tests until all desired functionality is implemented. This approach has been found to reduce defects and improve code quality \cite{sanchez2007sustained, maximilien2003assessing,nagappan2008realizing}.

Developers in exploratory programming ~\cite{kery2017exploring, trenouth1991survey} use edit-run cycles to rapidly prototype ideas and alternative solutions \cite{kery2017exploring,brandt2008opportunistic}. Data scientists often use exploratory programming~\cite{tukey1977exploratory}, where they may not know what the final output should be and use edit-run cycles to explore possible outputs. Researchers have built tools within computational notebooks to help data scientists explore alternative solutions faster by allowing them to navigate edit history~\cite{kery2017variolite, head2019managing} and interactively work with each running cell's output~ \cite{drosos2020wrex, lau2021tweakit, kery2020mage}.


Live programming tools offer direct support for edit-run cycles \cite{goldberg1983smalltalk,tanimoto1990viva, Tanimoto,mcdirmid2007living,lau2021tweakit, burckhardt2013it}. With live programming tools, developers edit their program and the tool automatically compiles, runs, and present the output~\cite{tanimoto1990viva, mcdirmid2016promise}. By automating the running step, live programming tools aim to make problem-solving and exploration more creative, programmers more productive, and programming, in general, more accessible \cite{rein2018exploratory}. Some live programming tools further support the edit-run cycle by generating code snippets that developers may use in the edit step. These tools use programming synthesis techniques, working from examples provided by developers or output direct manipulation to generate code snippets \cite{chugh2016programmatic,ferdowsifard2020small, santolucito2019live}. Lab studies of live programming tools have found that these tools can help developers comprehend and debug code more effectively \cite{lieber2014addressing,oney2014interstate}.

McDirmid \cite{mcdirmid2016promise} discusses ``the promise of live programming'', presenting the goal of creating environments in which the ``computer can then better assist the programmer in a fluid experience rich in feedback and affordance''. Much of the focus of these environments is on offering fast feedback and fluid transitions between the edit and run steps. However, little is known about how close developers' edit-run cycles are from a fluid ideal.


Prior work has designed a variety of tools enabling increased fluidity within edit-run cycles. Building on this work, we instead focus on investigating the edit-run behavior though analyzing and the characterizing edit-run cycles themselves. We contribute the first study of edit-run behavior in field setting. Our work informs tool designers about the nature of edit-run cycles and the barriers within these cycles to achieving fluidity.

\begin{table*}[]
\centering
\caption{The live-streamed programming videos used in the dataset.}
\label{tab:my-table}
\begin{tabular}{cccllcc}
\midrule
\multicolumn{2}{c}{Developer} & \multirow{2}{*}{\begin{tabular}[c]{@{}c@{}}\\ Observed Time\end{tabular}} & \multicolumn{3}{c}{Project} \\ \cline{1-2} \cline{4-7} 
   ID & Yrs. Exp.{\footnotesize{$^*$}}&& Name and Brief Description     & LOC & GitHub Stars & Commits  \\ \toprule
D1   & 10 & 2:00:56 & \textit{Firefox}: A popular web browser.     & 4M JavaScript   &2.1K & 751K
\\
D2   & 31 & 2:40:48 & \textit{Curl}: A library for transferring data.                & 138K C   & 19.9K &29K     
\\
D3 &
  7 &
2:46:10 &
  \textit{Serenity OS}:  A Unix-like operating system. &
 120K C++ & 9.8K & 18K
  \\
D4   & 8  & 3:28:35 & \textit{Tox}:  A library for task automation.                  & 10.7K Python & 2.1K & 2K  
\\
D5   & 8  & 2:59:49 & \textit{Downshift}:  A set of web components built with React. & 11K JavaScript& 9.4K& 664
\\
D6 &
  8 &
  2:23:48  &
  \textit{Uzual}:  A mobile app that helps track daily habits &
  2.5K JavaScript & 63 & 204
  \\

D7   & 10 & 2:54:43 & \textit{Vectrexy}: A game emulator.                            & 9.7K C++    & 39  & 656
\\
D8 &
  9 &
  3:05:25 &
  \textit{Kap}: A screen recorder for computers. &
  9.5K JavaScript & 13.5K & 870
  \\
D9   & 9  & 2:15:26 & \textit{DevBette}:  A web application for a small business.    & 3.6K C\# & 60  & 316      
\\
D10 &
  10 &
  2:17:14 &
  \textit{Alacritty}:  A cross-platform terminal. &
  16.5K Rust & 30.6K & 1.8K
  \\
D11  & 8  & 3:40:00 & \textit{Webpack}:  A bundler for javascript.                   & 95K JavaScript  & 57.8K & 13.3K
\\ 
        \bottomrule
      \multicolumn{6}{c}{\footnotesize{$^*$ \# of years contributing to open source projects.}}
\end{tabular}
\label{tab:dataset}
\end{table*}

\begin{table}[]
\centering
\caption{Definitions of the activities coded in the dataset.}
 \label{tab:activites}
\begin{tabular}{ll}
\midrule
Activity   & Definition            \\ \toprule
Browsing A File  & Open a file of code and leave it without edits.                   
\\
Editing A File  & Open a file of code and change its contents.         
\\
Testing Program & Run a program to observe its final output.                      
\\
Inspecting Program & Run a program to observe program state values.                   
\\
Consulting  Resources & Use the browser or local doc to search for info. 

\\
Other        
& Engage in noncoding work (e.g., writing notes). 
\\          
\bottomrule   
\end{tabular}
\end{table}

\section{Method} \label{Methods}
To answer our research questions, we began with a publicly available dataset of 28 hours of professional developer development work, annotated with activity codes. From this, we first identified edit and run steps by mapping activities to either edit or run steps. 
We then analyzed the sequence of steps to construct edit-run cycles within programming and debugging. Finally, we analyzed the activities to identify causes of gaps within and between edit-run cycles. 


\subsection{Dataset}
 We began our analysis with a dataset of live-streamed programming videos \cite{Alaboudi2019}, publicly available through the observe-dev.online platform\footnote{https://bit.ly/3kkbL2W}. The platform supports analysis of developer activity annotations alongside the recorded videos. The dataset and platform are described in detail in Alaboudi et al.~\cite{alaboudi2021exploratory}. We briefly review the dataset here.
\begin{figure}
    \centering
    \includegraphics[scale=.13]{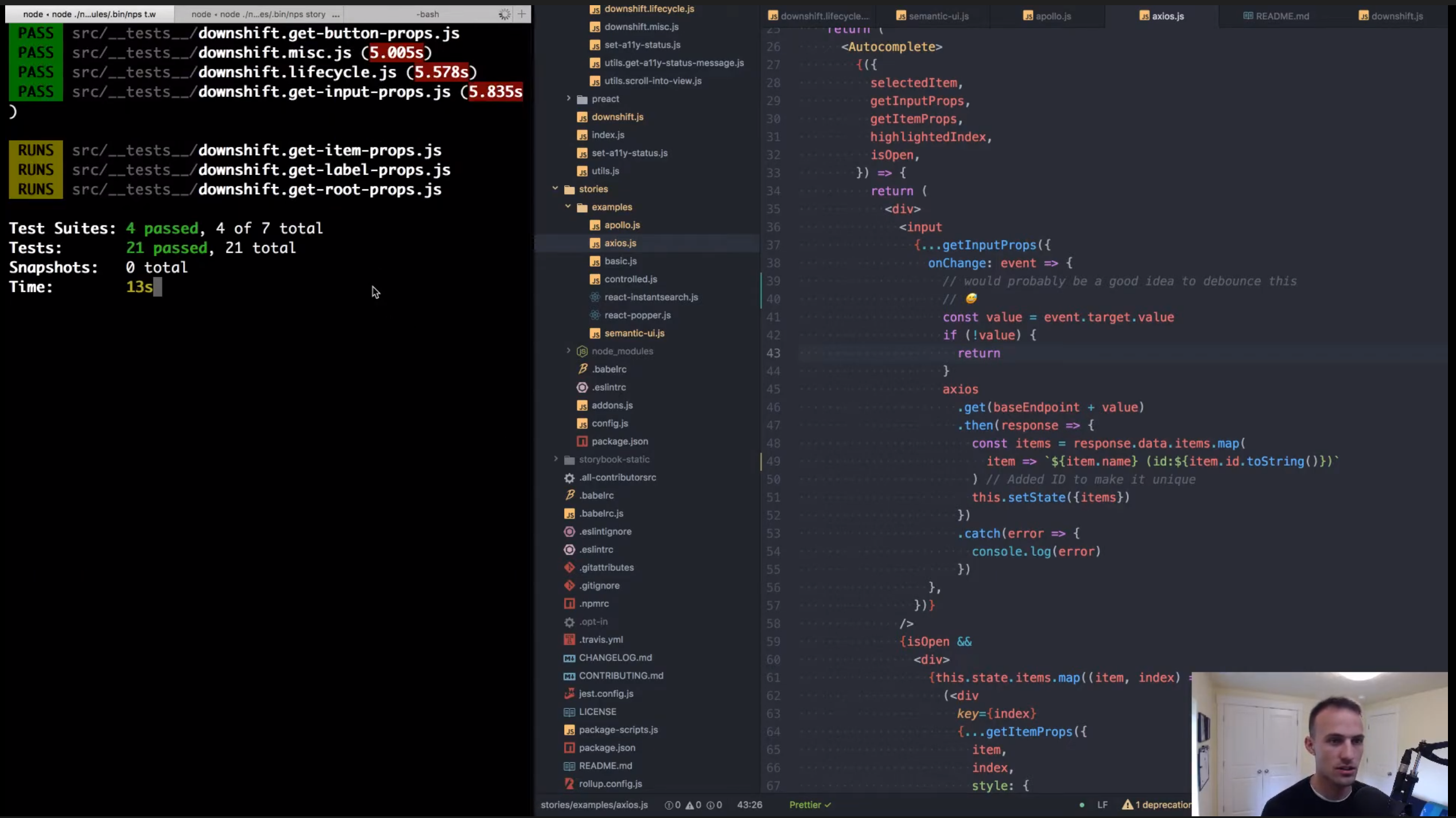}
    \caption{Live-streamed programming videos depict developers at work contributing to open source projects.}
    \label{fig:exampleLive}
\end{figure}
    \subsubsection{Live-streamed programming}
    Developers live-stream their development work on open source projects and invite other developers to join and watch \cite{Koebler2015}. The initial stream is then archived as a video on platforms such as YouTube and Twitch, making it available to other developers to watch asynchronously. What distinguishes these videos from other developers' typical screencasts \cite{MacLeod2015,ellmann2017find}  is that these videos are not created as a tutorial or documentation for particular API usage. Instead, developers show the entire development workflow as they work within open source projects~\cite{Alaboudi2019} (Figure \ref{fig:exampleLive}). 
    
    The dataset we selected contains live-streamed programming videos from 11 professional developers as they add new functionality and fix defects within open source projects. The 11 professional developers have been active contributors to open source projects for at least seven years. Many of these developers shared that they have worked for large companies such as Google, Microsoft, Lyft, PayPal, and Mozilla. The open source projects that developers contribute to within these videos vary in size, domain, and technologies. Projects are active, with hundreds of commits, tens of GitHub stars and forks, and thousands of lines of code. All projects also have a final executable version of the software for public use. Table \ref{tab:dataset} summarizes the characteristics of the annotated live-streamed programming videos included in the dataset.

\subsubsection{Developers activities}
\begin{figure}
    \centering
    \includegraphics[scale=.49, trim= .3cm 0cm .3cm 1.7cm, clip]{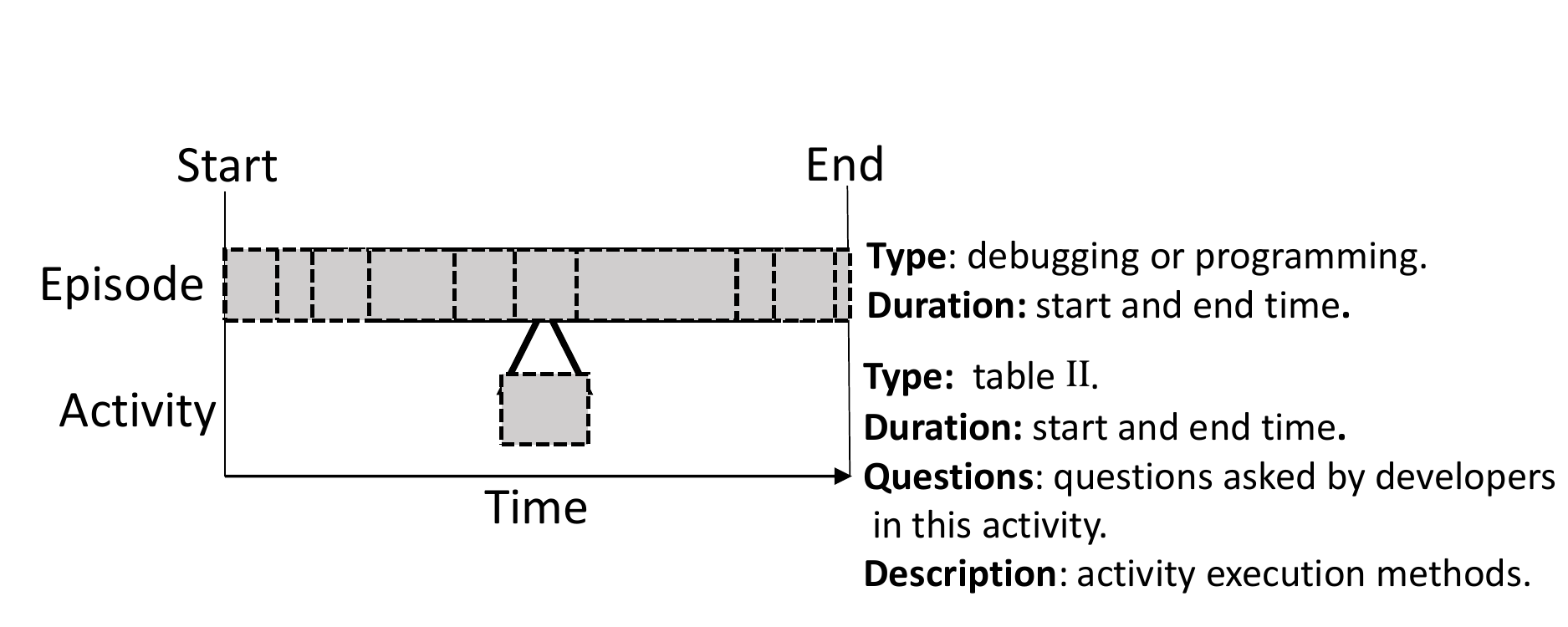}
    \caption{Development work is segmented into debugging and programming episodes. Each episode is composed of a sequence of activities. Each activity instance is annotated with several activity-specific codes describing how developers worked.}
    \label{fig:codedAnalysis}
\end{figure}

The dataset contains annotations of 3,544 activities that occurred in the videos, describing developers' moment-to-moment behavior while programming and debugging. Table \ref{tab:activites} summarizes the activity annotations in the dataset. Videos are first segmented into 89 debugging and 104 programming episodes. A debugging episode begins when a developer encounters a defect and begins work to fix it and ends when the developers either fixes the defect or stops debugging. A programming episode begins when developers do any kind of development work other than debugging, and ends when developers begin debugging or the video ends. Each episode consists of a sequence of activities (Figure \ref{fig:codedAnalysis}). For each activity instance, the dataset describes the start and end times, the questions developers asked during this time, and several activity-specific codes describing how work occurred. These include describing whether the developer tested the program manually or through automated tests for the test the program activity, inspected the program through the debugger or logs for the inspecting the program activity, and whether the developer used documentation or Q\&A forums for the consult resources activity.

\begin{figure}
    \centering
    \includegraphics[scale=.5, trim= 2.5cm 1.5cm 2cm 2cm, clip]{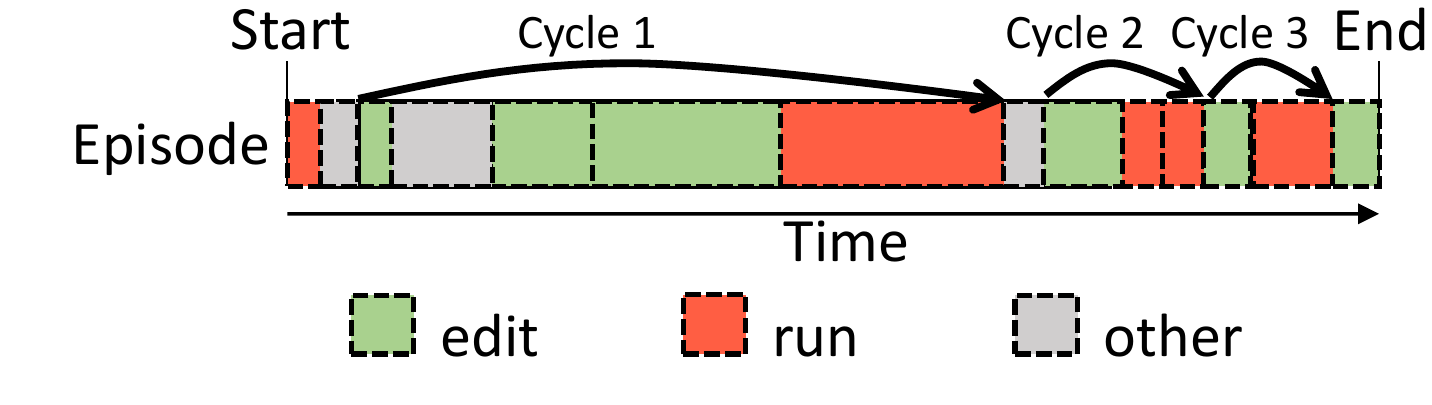}
    \caption{We first mapped browsing a file and editing a file activities to edit steps (green) and mapped testing the program and inspecting the program to the  run (red) steps. Edit-run cycles begin with the first edit activity and end with the last run activity. Run steps preceding the first edit (first activity) and edit steps following the last run step (last activity) do not belong to an edit-run cycle. Activities not mapped to either edit or run steps are included in the ``other'' category (gray), which may occur both within and between cycles. }
    \label{fig:edit-run cycles}
\end{figure}

\subsection{Data Analysis}
To analyze the coded videos, we constructed and analyzed edit-run cycles. To our knowledge, edit and run steps within live programming have not before been precisely defined. Therefore, we inferred definitions of these steps from tools built to support edit-run cycles~\cite{rein2018exploratory}. We define the edit step as the set of activities in which developers browse, comprehend, and edit code. We define the run step as the set of activities in which developers run the program to test for the correct output or inspect intermediate states created by executing the program (e.g, stepping through the program using the debugger).  Using these definitions, we mapped activity browsing a file or editing a file to the edit step. We then mapped activity testing or inspecting a program to the run step. All other activity was mapped to other types of work. Each cycle begins with an edit step activity and ends with a run step activity. Other activities may occur within or between cycles and constitute gaps within and between edit-run cycles. Figure \ref{fig:edit-run cycles} depicts the sequencing of activities within edit-run cycles.

There were a small number of incomplete cycles with only an edit or run step. These  occurred at the beginning or end of episodes. We dropped incomplete cycles as they represent duplicate cycles in the subsequent episode. For example, developers may start a debugging episode with a run step that was also part of a cycle in the previous programming episode. In this example, we count only the complete cycle in the programming episode and drop the run step at the beginning of the debugging episode as a duplicate.

We wrote an automated script to map each activity instance to an edit or run step. To verify the correctness and completeness of the automated script, we  manually coded cycles for a random sample from each developer. We then ran the script and identified incorrect cases. After iteratively revising the scripts, the manually coded cycles exactly matched the cycles generated by the scripts.

To compare the edit-run cycles within debugging and programming work, we compared edit-run cycles within debugging and programming episodes. Our analysis revealed that the episodes in debugging (n = 89, $\mu$ = 10 minutes, median = 4 minutes) and programming (n = 104, $\mu$ = 8 minutes, median = 4 minutes) have the same duration distribution (two-sample Kolmogorov-Smirnov p-value = 0.68, D = 0.1), making them suitable to divide work into programming and debugging with comparable units of time. 

Finally, to better understand the causes of developers interrupting edit-run cycles to do other work, we analyzed each cycle, as well as transition between cycles containing other work. We analyzed their behavior by examining the questions they asked as well as the videos. In many cases, we needed to observe a developer's work beyond an edit-run cycle to understand their intention. The dataset of edit-run cycles and our analysis are available in the replication package\footnote{https://bit.ly/3r1QDSD}.

\section{results}
Our analysis yielded 581 and 207 edit-run cycles for debugging and programming, respectively. Table \ref{tab:studyData} lists the number and length of the edit-run cycles we examined. Due to the skewness of the data, we report the variance using interquartile ranges and Letter-Value plots. Letter-Value plots encode more information about skewed distributions than traditional boxplots, recursively subdividing the remaining region in half based on its median (e.g., fourths, eights, sixteenths) \cite{hofmann2017valu}.

 
\begin{table}[]
\centering
\caption{The number and total time of episodes and edit-run cycles.}
 \label{tab:studyData}
 \resizebox{\linewidth}{!}{%

\begin{tabular}{lcccc}
\midrule
Episode Type   &  \# of Episodes &  Total Episode Time &  \# of Cycles &  Total Cycle Time            
\\ \toprule
Debugging& 89 & 15.3 hours  & 581&14.4 hours\\    
Programming& 104& 13.1 hours  &207& 11 hours \\
\bottomrule   
\end{tabular}
}
\end{table}

\begin{figure*}
\begin{subfigure}{\textwidth}
  \centering
        \includegraphics[scale=.5,trim= 1.5cm 8.6cm 0cm 3cm, clip]{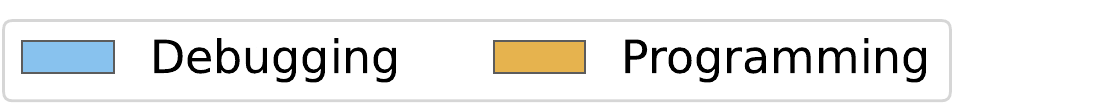}

        \label{fig:cycleDuration_Density}
\end{subfigure}
\begin{subfigure}{.51\textwidth}
  \centering
        \includegraphics[scale=.55, trim= 2cm 0cm 0cm .7cm ]{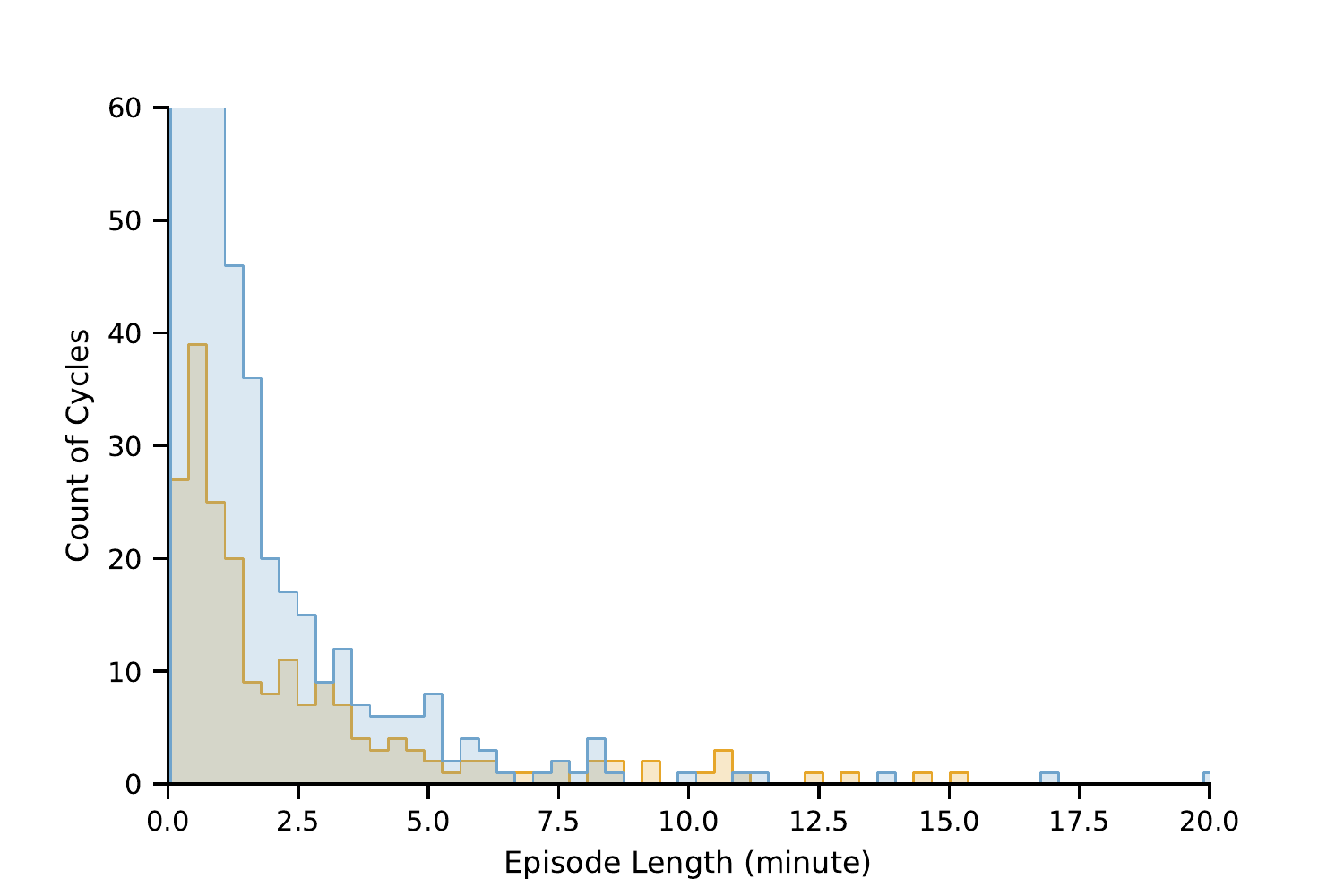}
        \caption{The distribution of edit-run cycle length in debugging and programming.}
        \label{fig:cycleDuration_Density}
\end{subfigure}
\begin{subfigure}{.51\textwidth}
  \centering
        \includegraphics[scale=.4, trim= 5cm .2cm 2cm 0cm ]{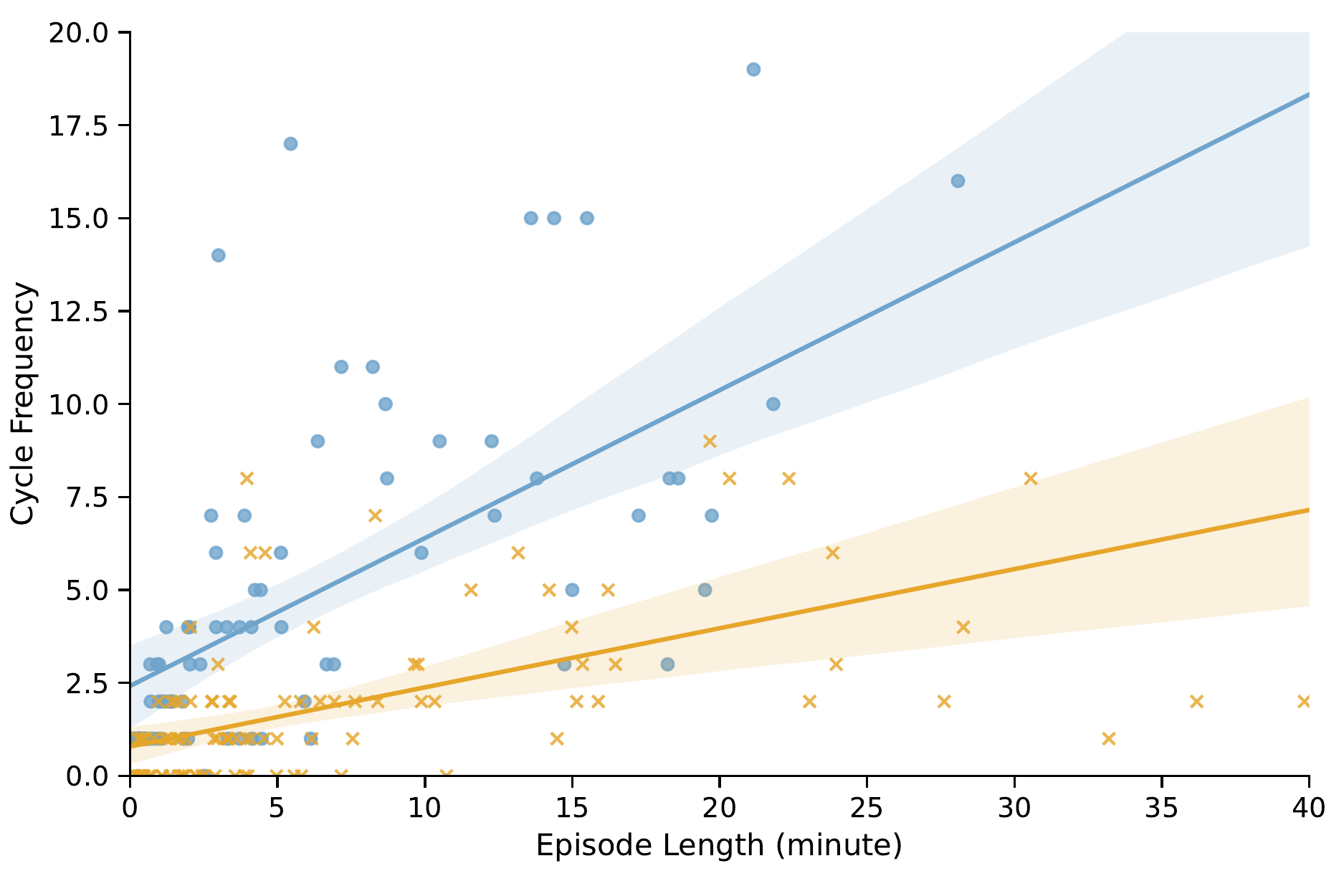}
        \caption{The relationship between cycle frequency and episode length in debugging and programming.}
  \label{fig:context2}
\end{subfigure}

\begin{subfigure}{.51\textwidth}
  \centering
  \includegraphics[scale=.35, trim= 1.5cm 0cm 0cm 0cm clip]{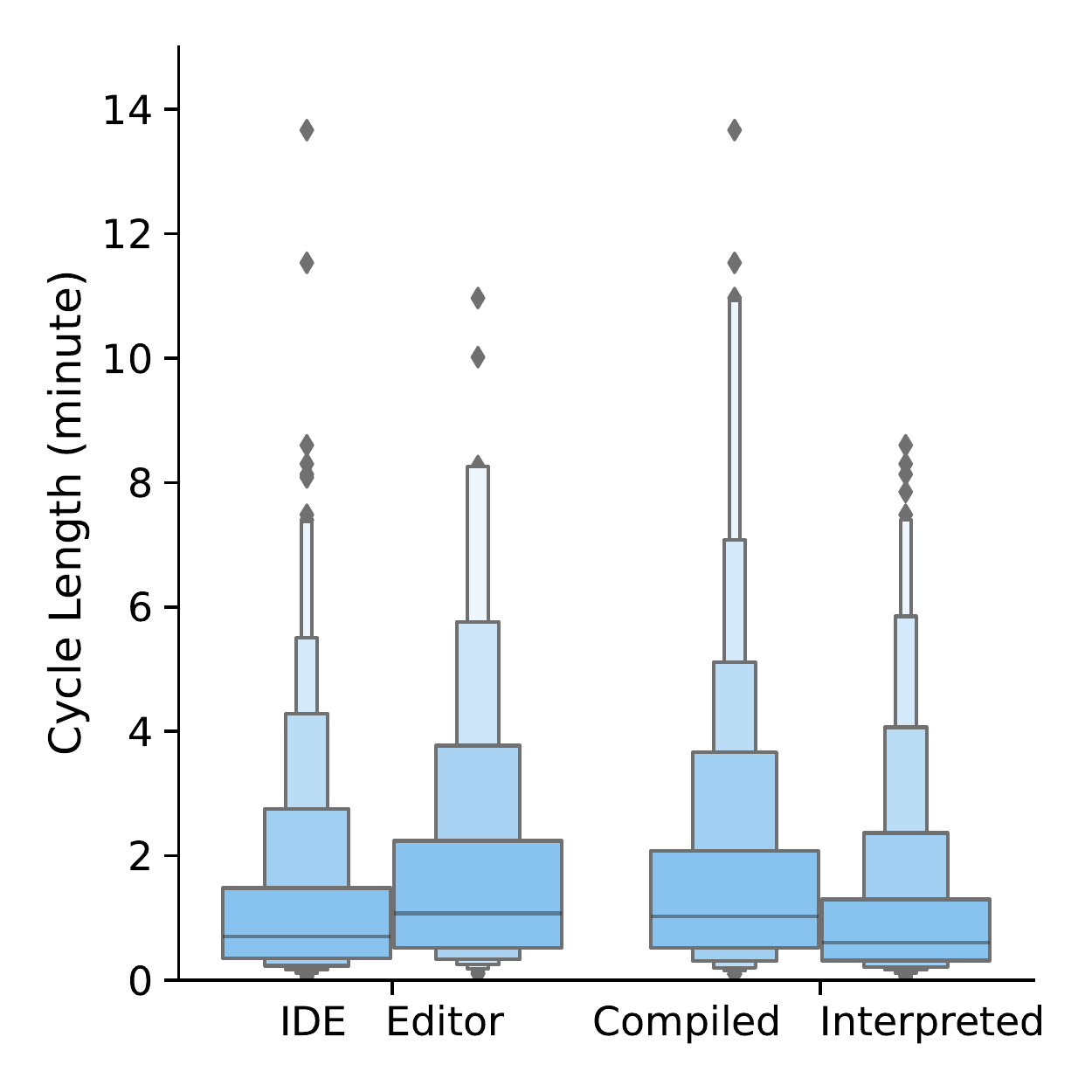}
  \includegraphics[scale=.35, trim= 1.8cm 0cm 0cm 0cm  clip]{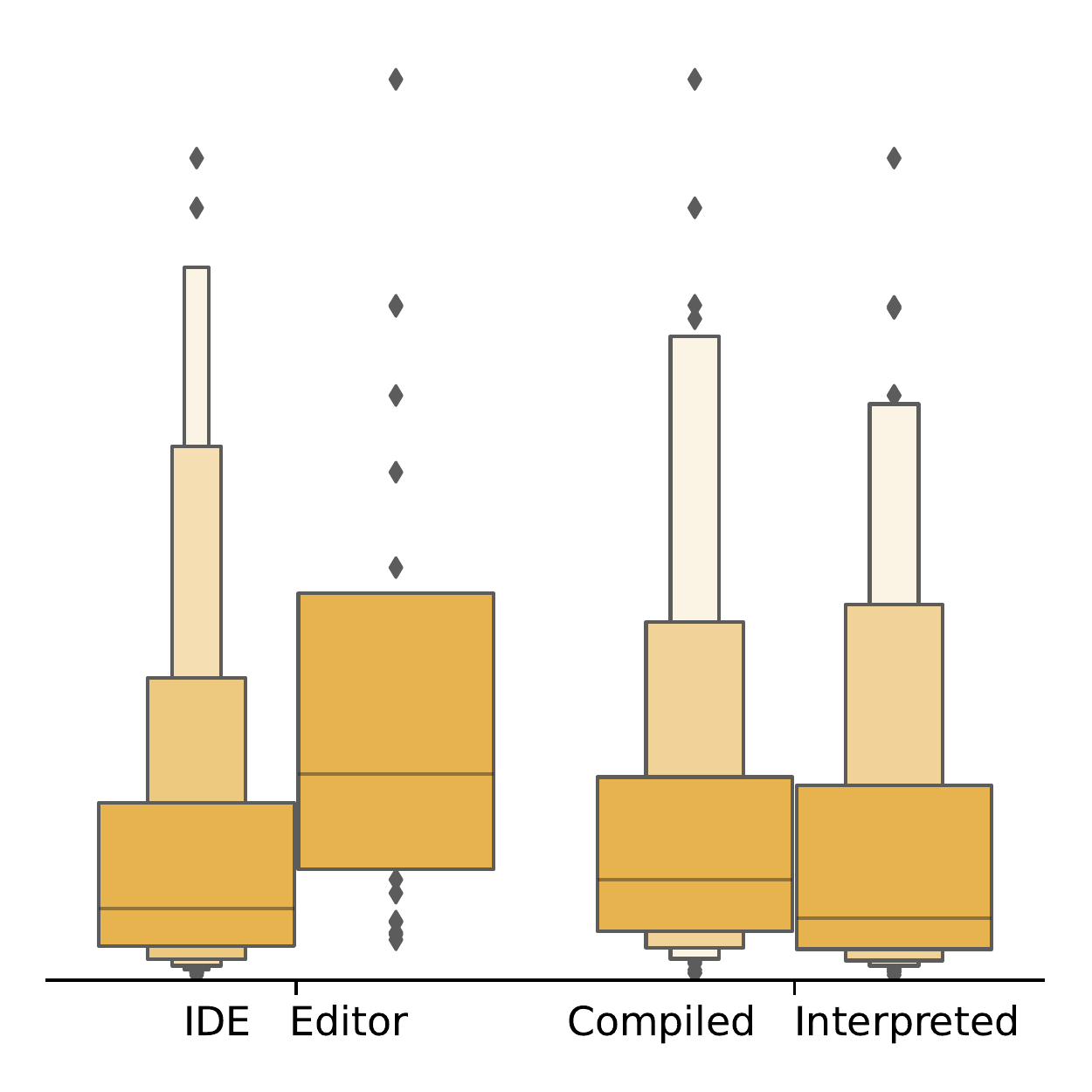}

  \caption{A Letter-Value plot of cycle length by the type of development environment and programming language.}
  \label{fig:context3}
\end{subfigure}
\begin{subfigure}{.51\textwidth}
  \centering
  \includegraphics[scale=.35, trim= 1.5cm 0cm 0cm 0cm clip]{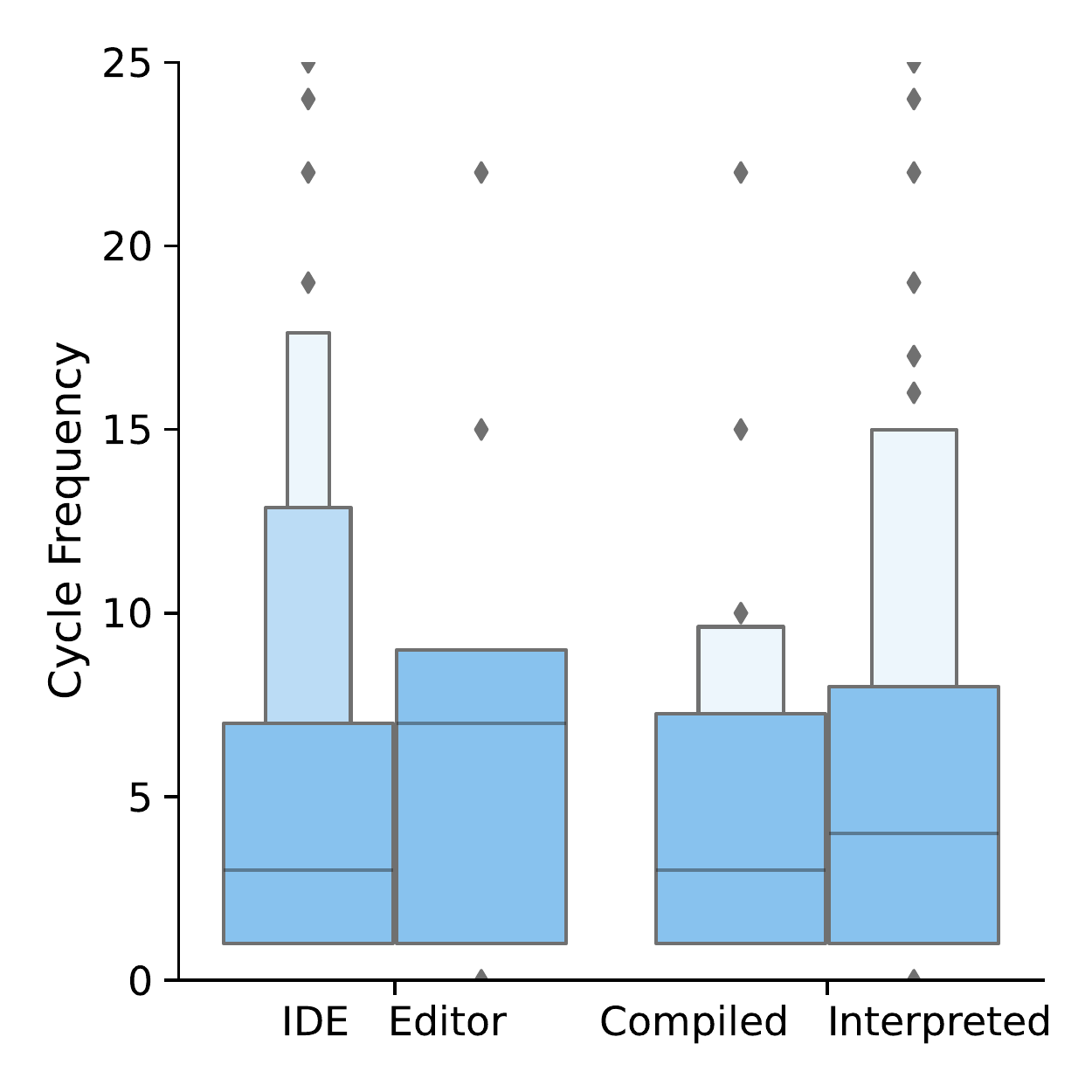}
  \includegraphics[scale=.35, trim= 1.7cm 0cm 0cm 0cm  clip]{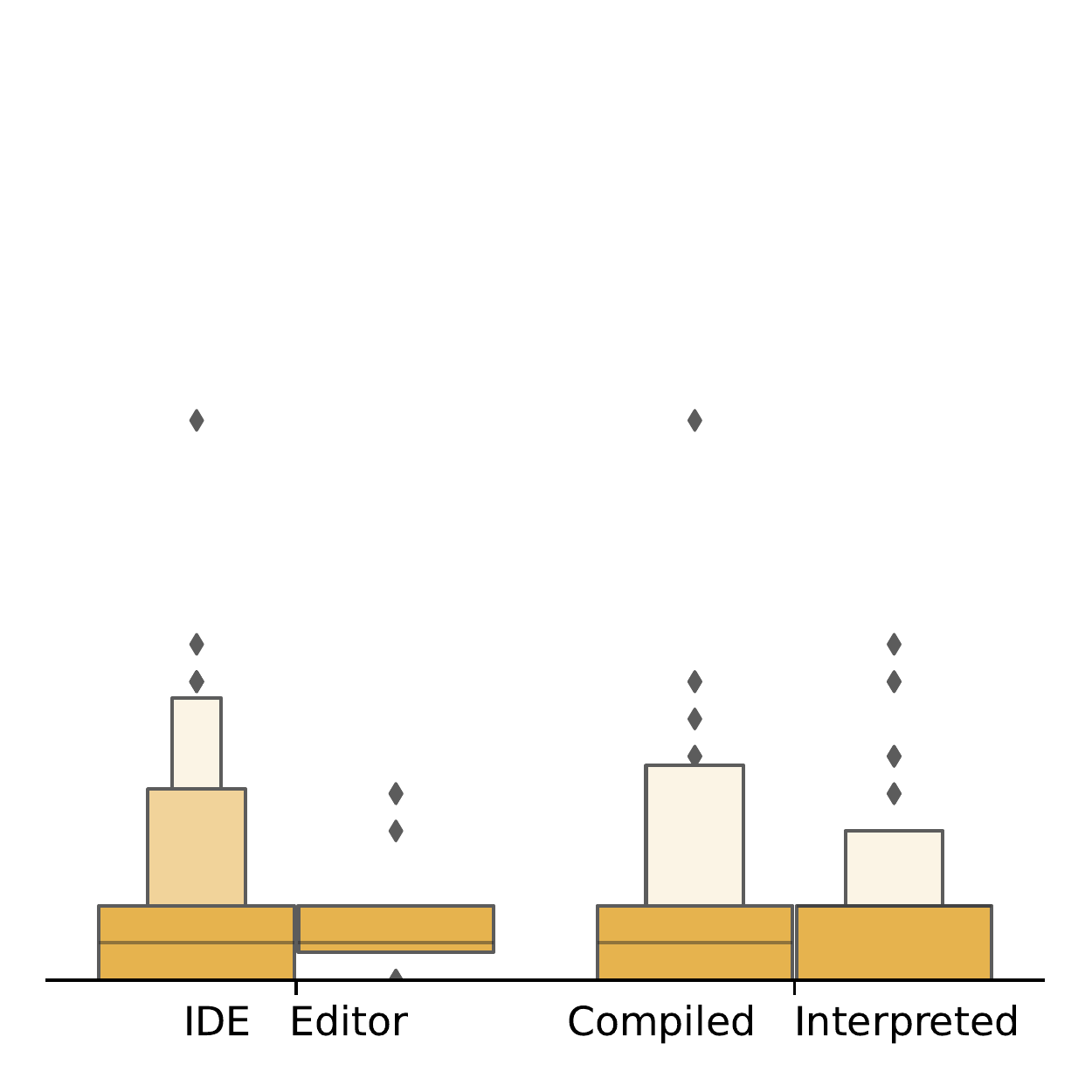}

  \caption{
  A Letter-Value plot of cycle frequency by the type of development environment and programming language.}
  \label{fig:context4}
\end{subfigure}

\caption{The distribution of edit-run cycle length and frequency (cycles per episode).}
\label{fig:context}
\end{figure*}

\subsection{Length and Frequency of Edit-Run Cycles}
A key property of the fluid interaction envisioned by live-programming are short and frequent edit-run cycles.
We found that most edit-run cycles within both debugging and programming work were less than five minutes. Table \ref{tab:q1Table} lists the mean count of cycles per episode (cycle frequency) and cycle length.
 
\begin{table}[]
\centering
\caption{The mean episode length, cycle frequency, and cycle length. The interquartile range is reported in parentheses (Q\textsubscript{1}-Q\textsubscript{3}).}
\label{tab:q1Table}
 \resizebox{\linewidth}{!}{%

\begin{tabular}{lccc}
\midrule 
Episode
& 
Episode length
&
Cycles Frequency
& 
Cycle Length

\\
 \toprule
Debugging   
&
 10 (1-12) minutes 
&
 7 (1-9)
&
1 (0.4-2) minutes

\\
Programming  
&
\hspace{.05cm} 8 (2-10) minutes
&
 2 (0-2)
&
3 (1-3)\hspace{.21cm} minutes
\\
\bottomrule
\end{tabular}
 }
\end{table}



Edit-run cycles in programming tended to be longer than edit-run cycles in debugging. On average, edit-run cycles in programming episodes were 3 minutes, with only 15\% longer than five minutes. Edit-run cycles in debugging were one minute, with only 5\% longer than five minutes. Figure \ref{fig:cycleDuration_Density} plots the distribution of cycle length for debugging and programming.


Debugging episodes contained an average of 7 edit-run cycles per episode, as developers edited and ran the program 7 times before ultimately fixing the defect. Programming episodes contained an average of 2 edit-run cycles per episode, as developers tended to find the output not as expected and begin debugging after running the program twice. We found a strong positive correlation between debugging episode length and cycle counts (Pearson r = 0.88), indicating developers continue running the program regularly as debugging episodes grow in length. However, there was only a moderate positive correlation between programming episode length and cycle counts (Pearson r = 0.59) (Figure \ref{fig:context2}). This indicates that, as programming episodes grow in length, developers do not continue to regularly run the program.
\begin{table}[]
\centering
\caption{Mean cycle frequency (cycles per episode).}
 \resizebox{\linewidth}{!}{%

\label{tab:freqContext}
\begin{tabular}{lccccccc}
\midrule
&
  \multicolumn{2}{c}{Programming Language} & &
  
  \multicolumn{2}{c}{Development Env.} \\
\cline{2-3} \cline{5-6}
Episode
 &
  {Compiled} &
  {Interpreted} &
  {Difference} &
  {IDE} &
  {Editor} &
  {Difference} \\
  \toprule
  {Debugging} &
  {7.2} &
  {6} &
  {19\%} &
  {5.9} &
  {10} &
{69\%} \\
  {Programming} &
  {2.1} &
  {1.9} &
  {11\%} &
  {2} &
  {1.7} &
{21\%} \\
  
  \bottomrule
\end{tabular}%
}
\end{table} 
\begin{table}[]
\centering
\caption{Mean cycle length (seconds).}
 \resizebox{\linewidth}{!}{%

\label{tab:durationContext}
\begin{tabular}{lccccccc}
\midrule
&
  \multicolumn{2}{c}{Programming Language} & &
  
  \multicolumn{2}{c}{Development Env.} \\
\cline{2-3} \cline{5-6}
Episode
 &
  {Compiled} &
  {Interpreted} &
  {Difference} &
  {IDE} &
  {Editor} &
  {Difference} \\
  \toprule
  {Debugging} &
  {108.5} &
  {70.3} &
  {54\%} &
  {77.9} &
  {120.4} &
  {55\%} \\
  {Programming} &
  {185.7} &
  {169.4} &
  {10\%} &
  {155.9} &
  {314.3} &
  {102\%}
  \\
  \bottomrule
\end{tabular}%
}
\end{table}

We also examined how edit-run behavior varied with the type of programming language (compiled or interpreted) and development environment (IDE or traditional text editor) (Figure \ref{fig:context3}, Figure \ref{fig:context4}, Table  \ref{tab:freqContext}, and Table \ref{tab:durationContext}). 
Developers who used a text editor (D2, D4, D10) completed 69\% more edit-run cycles when debugging and spent 55\% and 102\% more time in each cycle when debugging and programming, respectively, than developers who used an IDE (D1, D3, D5-D9, D11). Developers who used compiled languages (D2, D3, D7, D9, and D10) completed 11\% more cycles with 10\% more time in each cycle when programming and 19\% more cycles with 54\% more time each when debugging than developers who used interpreted languages (D1, D4, D5, D6, D8, and D11).




\begin{table}[]
\centering
\caption{Characteristics of edit and run steps by episode type (debugging or programming). }
\label{tab:steps}
\resizebox{\linewidth}{!}{%
\begin{tabular}{p{4.5cm}p{4.5cm}}
\hline
Summary & Debugging (\includegraphics[scale=.2, trim =0cm 0cm 1cm 0cm, clip] {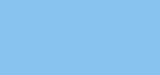})  vs.
Programming (\includegraphics[scale=.2, trim =0cm 0cm 1cm 0cm, clip] {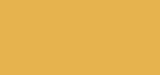})
\\  \toprule

The edit step lasted for one minute on average and was twice as long in programming as in debugging. The run step lasted for about half a minute for both programming and debugging. &
 {\multirow{2}{*}{\includegraphics[scale=.23,trim=1cm 0cm 0cm 0cm]{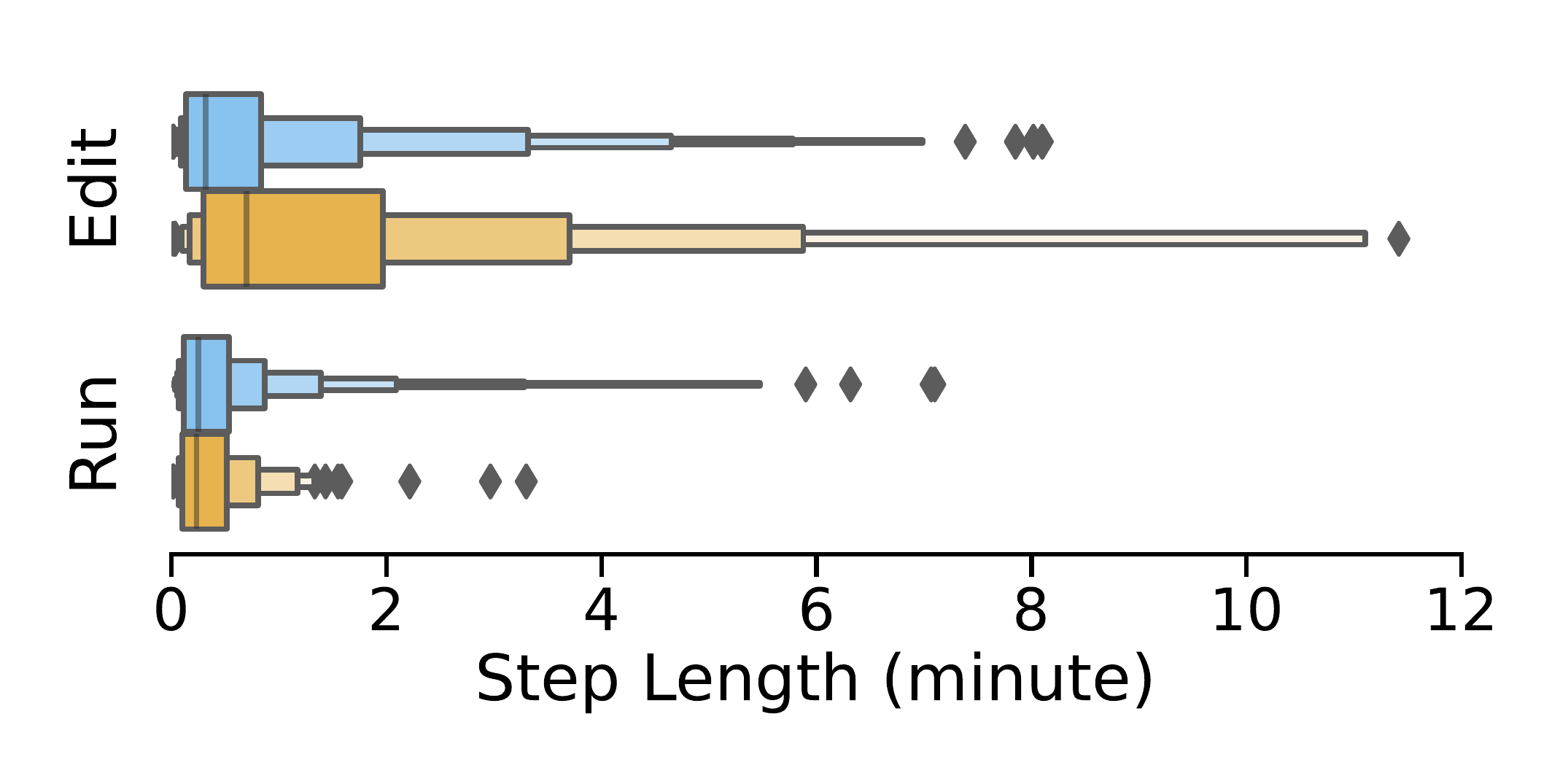}}}
  \vspace{2cm}\\
  Edit steps occupied the majority of edit-run cycle time in debugging and programming.&
 {\multirow{2}{*}{\includegraphics[scale=.24,trim=1cm 0cm 0cm .7cm]{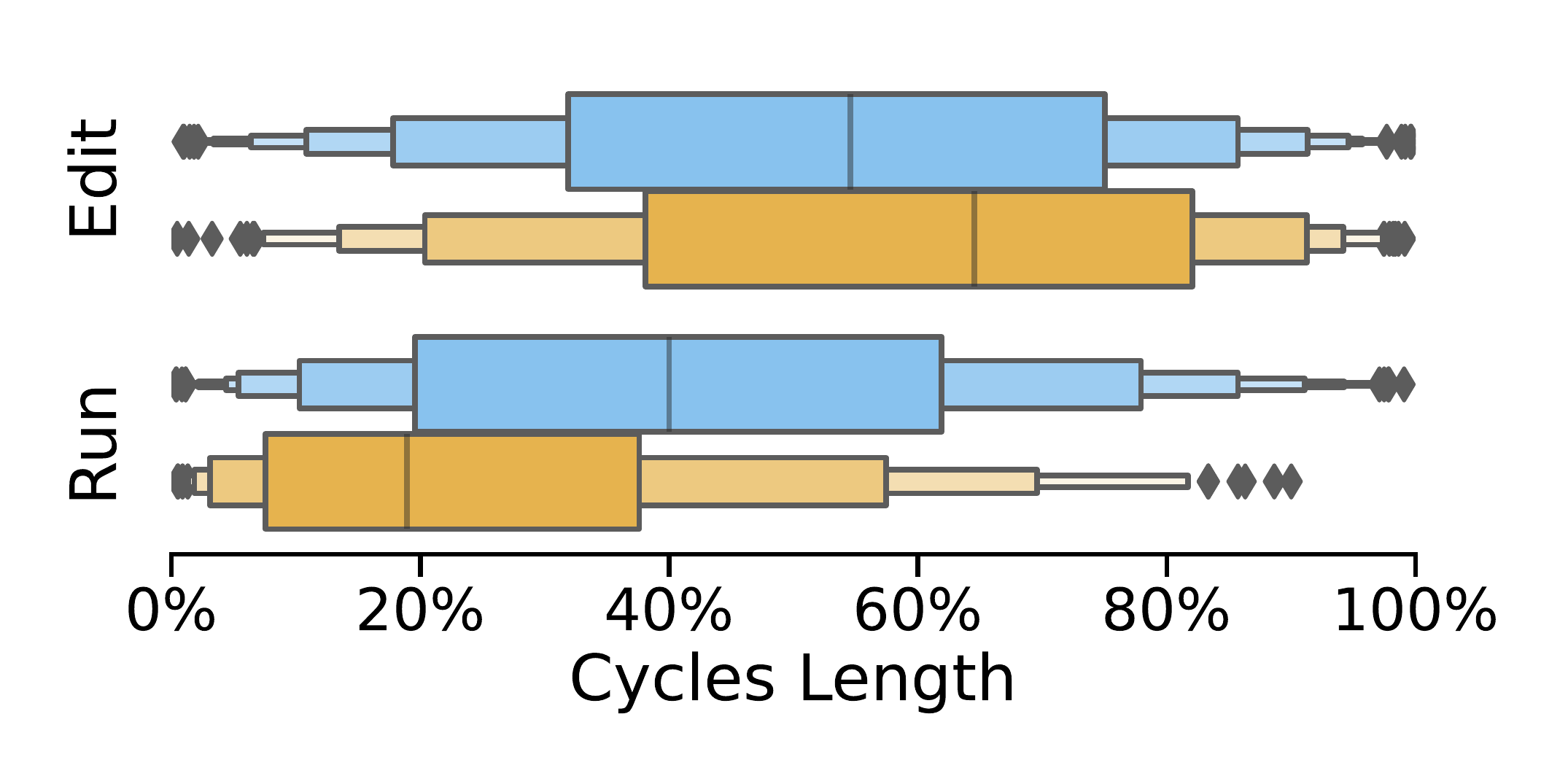}}}
  \vspace{2cm}\\
  During each edit step, developers on average edited only one file.&    {\multirow{2}{*}{\includegraphics[scale=.25, trim=1cm 0cm 0cm 1cm]{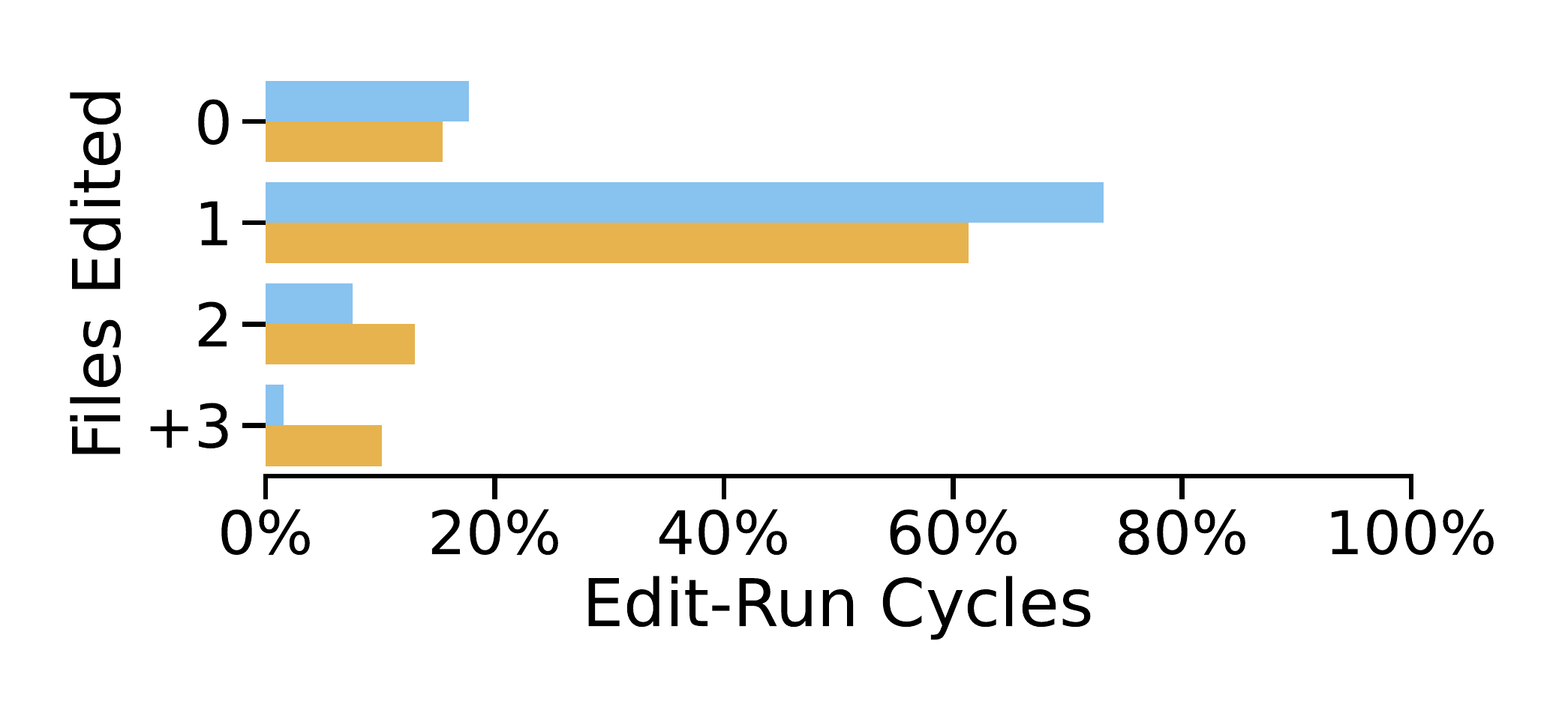}}}
   \vspace{1.5cm}\\
   During an edit step, developers usually edit files they visit. Developers visited files that they only browsed in only 25\% of the edit steps.&
  {\multirow{2}{*}{\includegraphics[scale=.25, trim=1cm 0cm 0cm 0cm]{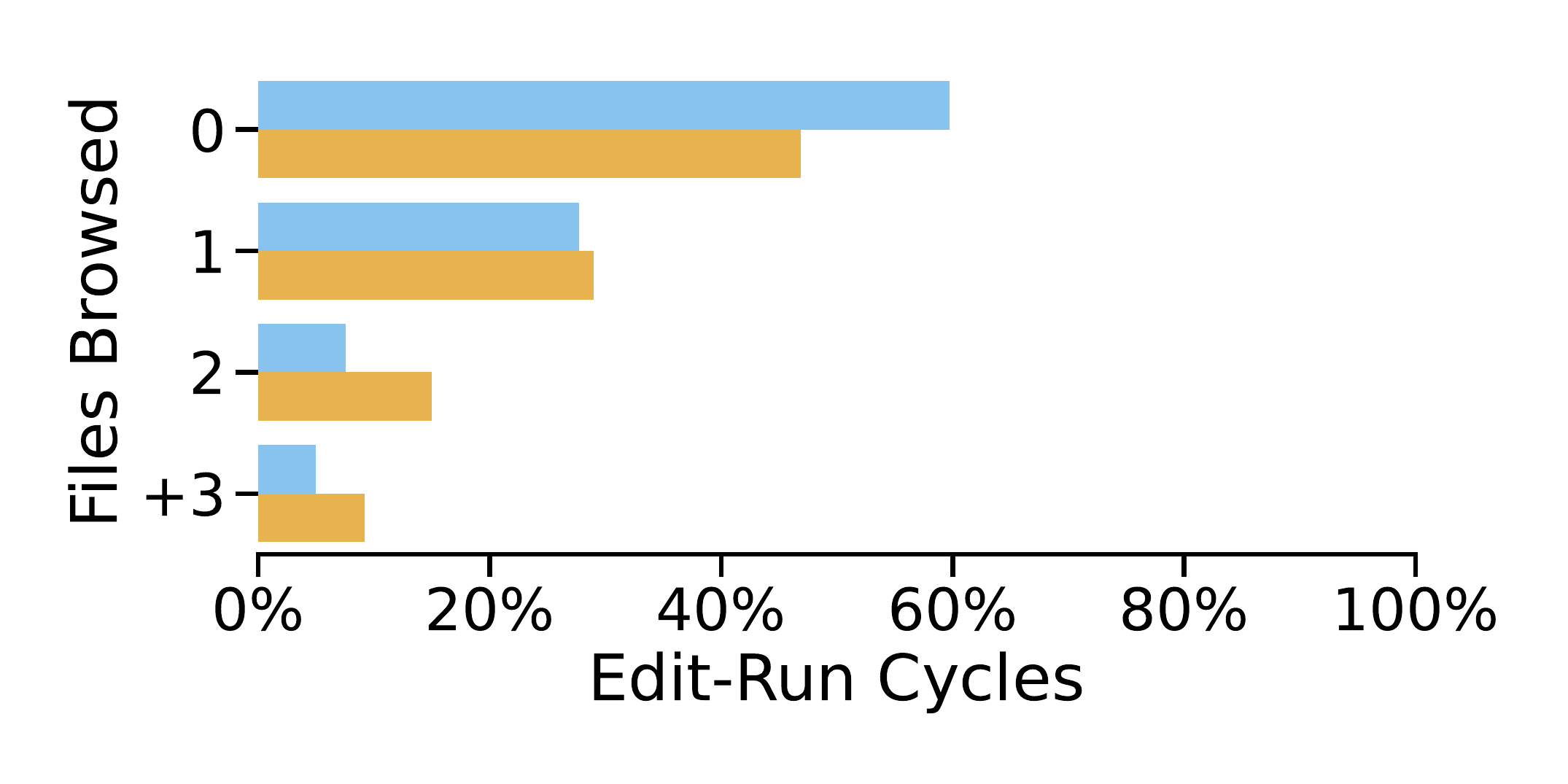}}}
\vspace{2cm}\\
  Developers usually run the program manually (manual). Only 18\% of run steps used automated tests (tests). &  {\multirow{2}{*}{\includegraphics[scale=.24, trim=1cm 0cm 0cm 0cm] {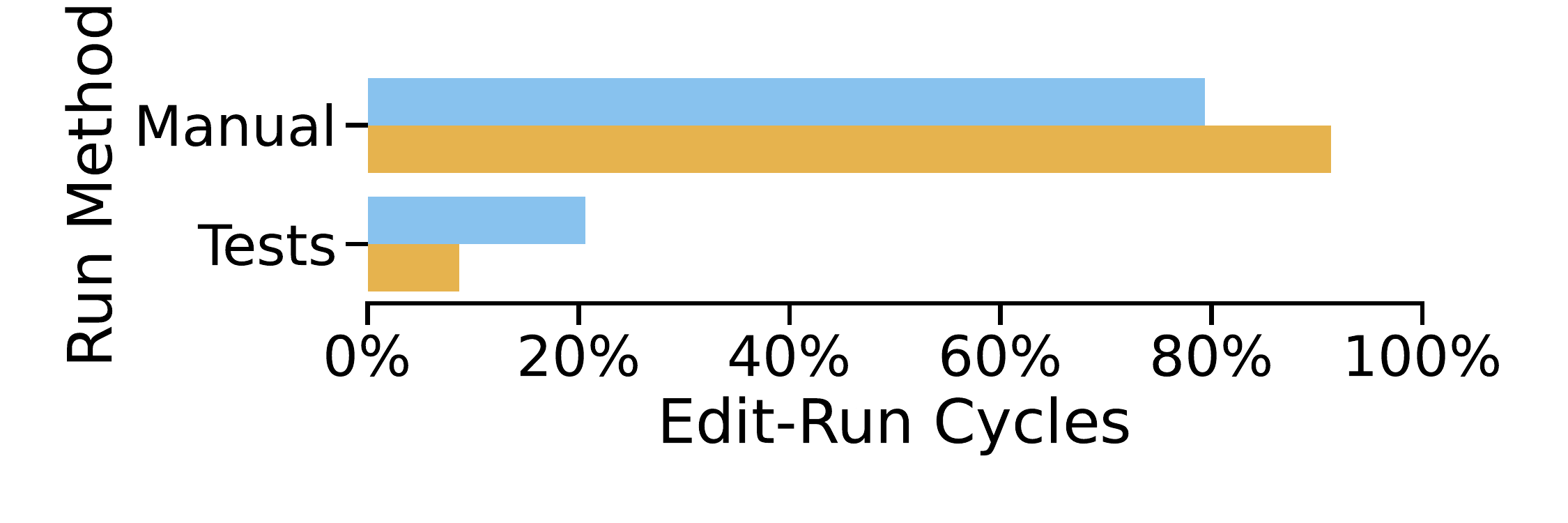}}}
  \vspace{1.5cm}\\
Developers usually run the program to observe the final output (final). Only 23\% of run steps used logs or the debugger (states).& 
  {\multirow{2}{*}{\includegraphics[scale=.24, trim=1cm 5cm 0cm 0cm] {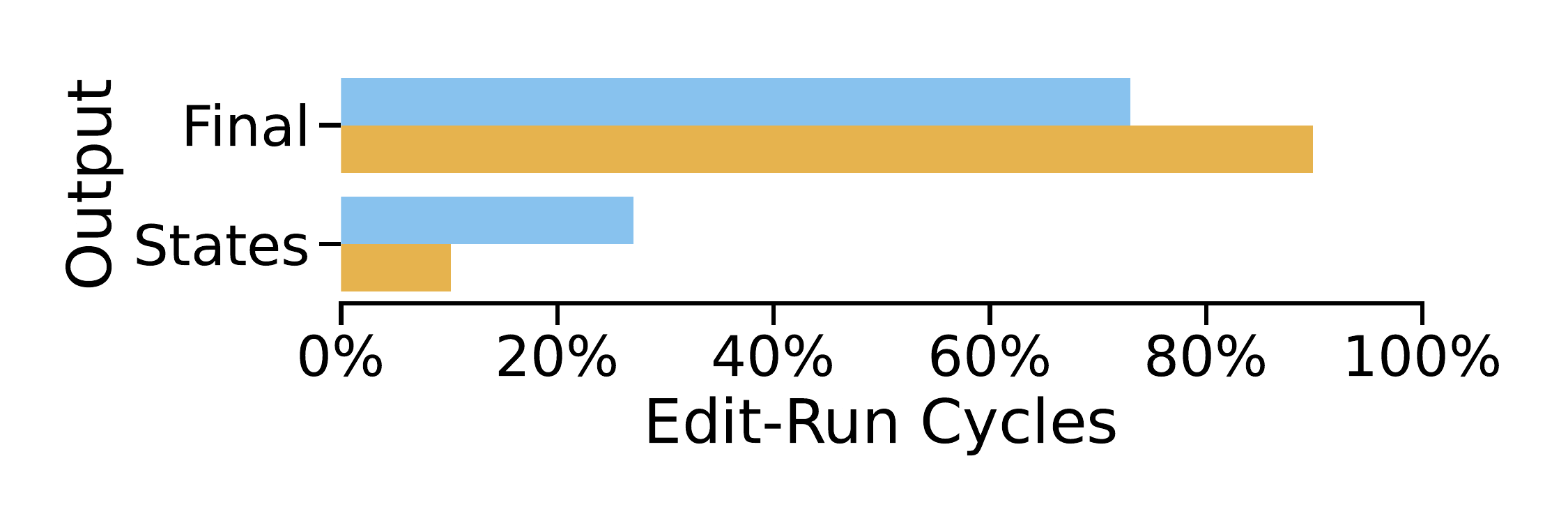}}}
   \vspace{1.5cm}\\
  \bottomrule

\end{tabular}
}
\end{table}

\subsection{Edit and Run Activities}
We mapped 2026 activities to an edit step and 828 activities to a run step, identifying edit and run steps within each of the 788 edit-run cycles. Table~\ref{tab:steps} summarizes the characteristics of edit and run steps within debugging and programming work.  

In live programming, the emphasis on fluidity is largely focused on enabling developers to remain focused on the edit step. We found that in both debugging and programming developers spent the majority of the edit-run cycle time editing. 
The edit step constituted on average 59\% ($\pm$27\%) of the edit-run cycle time in programming and 53\% ($\pm$26\%) in debugging. Run steps constituted less time, but still constituted much of the cycle length in debugging. On average, the run step consumed 26\% ($\pm$23\%) of cycle time in programming and 42\% ($\pm$26\%)  in debugging.     

Developers generally focused on editing a single file in programming and particularly when debugging. Only 20\% of edit-run cycles in programming involved editing more than one file and only 10\% of edit-run cycles when debugging. Edit-run cycles far more often involved browsing other files, with  40\% of debugging cycles and 53\% of programming cycles involving browsing one or more files. Note that the sum of the edit and run steps may not equal the total cycle time. This is due to gaps where developers did other work besides edit and run steps (Section \ref{otherWork}).

Developers generally executed the program manually through the GUI or command line rather than through automated tests. This was particularly true for programming. Only 9\% of edit-run cycles in programming and 21\% of edit-run cycles in debugging involved running the program through automated tests. Developers mostly ran the program to observe its final output. Developers only observed program state through logs or the debugger within 27\% and 10\% of the cycles.

\begin{figure*}
\begin{subfigure}{.5\textwidth}
  \centering
  \includegraphics[scale=.23, trim= 11.5cm 2cm 6cm 1cm,  clip]{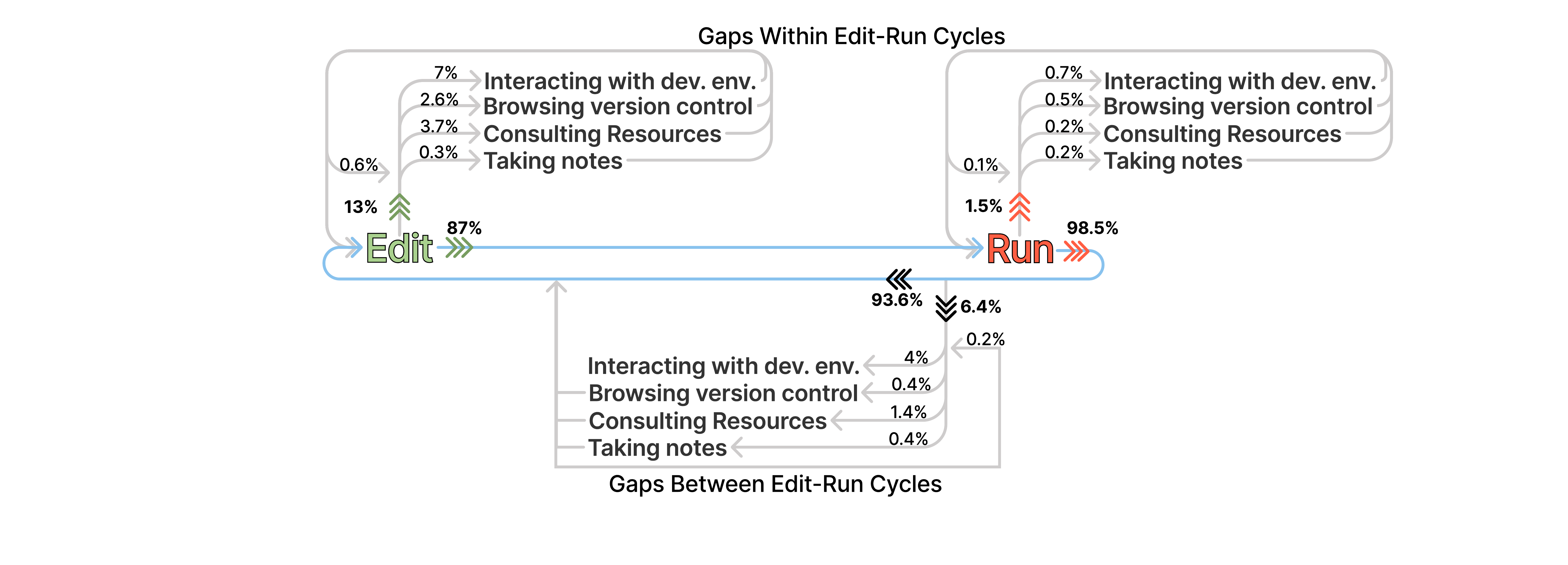}
  \caption{581 edit-run cycles in debugging episodes with 493 transitions.}
  \label{fig:sub-first}
\end{subfigure}
\begin{subfigure}{.5\textwidth}
  \centering
  \includegraphics[scale=.23, trim= 10cm 2cm 6cm 1cm,  clip]{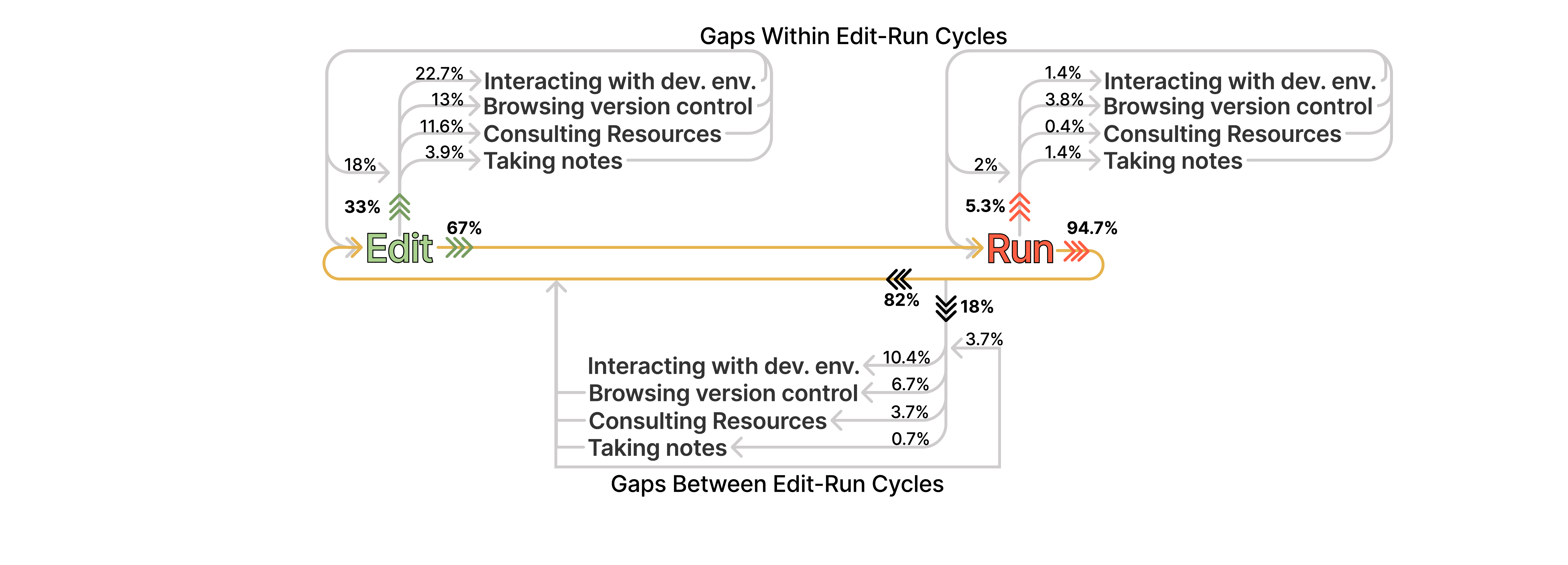}
  \caption{207 edit-run cycles in programming episodes with 193 transitions.}
  \label{fig:sub-second}
\end{subfigure}


\caption{Transitions between edit and run steps in (a) debugging and (b) programming, with gaps within cycles (top) and between cycles (bottom) where developers worked on other activities. }
\label{fig:other}
\end{figure*}

\subsection{Gaps Within and Between Edit-Run Cycles} \label{otherWork}
In perfectly fluid live programming, developers sequentially engage in edit-run cycles, beginning the next cycle without any interrupting other work between or within the cycle. 
We found that most cycles did not contain gaps during or between them.
94\% of edit-run cycles in debugging and 82\% of edit-run cycles in programming were sequential with no gaps in which developer engaged in activities beyond the edit or run steps. In some cases, gaps occurred, where developers engaged in other activities (Figure \ref{fig:other}). 
In the following sections, we examine the gaps within and between edit-run cycles as well as their causes.

\begin{figure}
        \centering
        \includegraphics[scale=.4, clip]{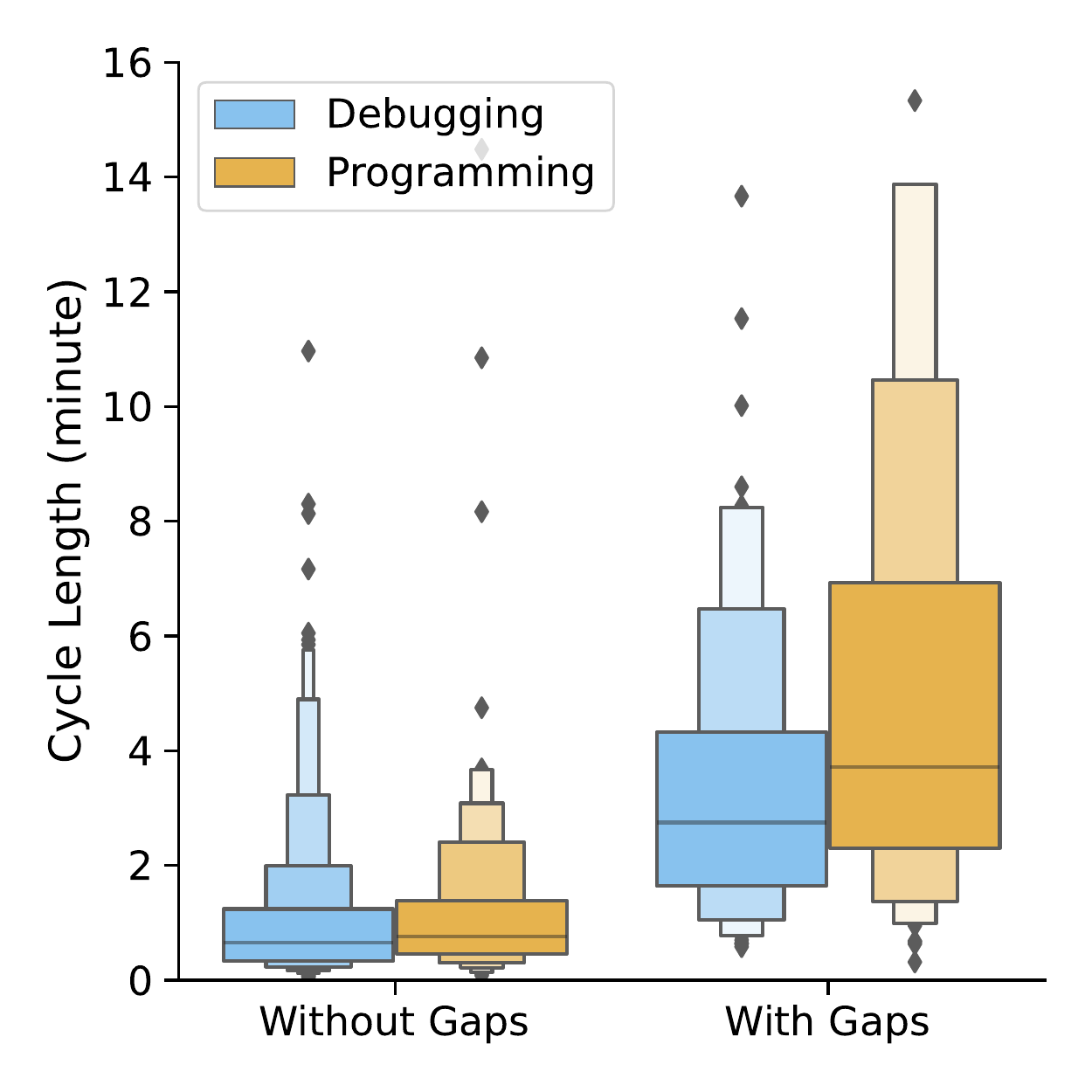}
        \caption{The distribution of cycle length without and with gaps.}
        \label{fig:otherWorkcycles}
    \end{figure}

\subsubsection{Gaps within edit-run cycles} 
Edit-run cycles contained gaps during the edit step much more often (33\% in programming, 13\% in debugging) than during the run step (5.3\% in programming, 1.5\% in debugging). During gaps in the edit step, developers most often interacted with their development environment (7.7\% of debugging and 22.7\% of programming edit-run cycles). This included using tools to search for code snippets across the codebase, installing third party libraries, and navigating between files. Consulting resources was also a common type of work during gaps in the edit step (3.7\% of debugging and 11.6\% of programming edit-run cycles). Browsing the issue tracker was much more common in edit step gaps in programming (13\%) than in debugging (2.6\%). Run steps did not contain gaps as often as edit steps, showing more focus and less need to do work beyond observing the program output. 

Gaps within edit-run cycles introduce friction, reducing the fluidity of developers interactions when editing code and observing the output. We found that edit-run cycles with gaps were four times longer (Figure \ref{fig:otherWorkcycles}). Edit-run cycles without gaps (n = 629) lasted an average of 1 minute. In contrast, edit-run cycles containing gaps (n = 159) lasted an average of 5 minutes.

\subsubsection{Gaps between edit-run cycles}
Developers most often transitioned from one edit-run cycle to the next in programming (n = 135) and debugging (n = 493) sequentially without any gaps. Transitions between debugging edit-run cycles had fewer gaps than programming edit-run cycles. 18\% of transitions between programming edit-run cycles contained gaps while only 6.4\% of transitions between debugging edit-run cycles contained gaps. When a transition contained gaps, developers spent on average 22 seconds when debugging and 53 seconds when programming. The most common type of other work developers did during the gaps between cycles wan interacting with the development environment (4\% debugging transitions and 10.4\% programming transitions).

\subsubsection{Causes of gaps within and between edit-run cycles}
To understand the cause of gaps which interrupted fluid edit-run cycles, we examined the work developers did during these gaps. We identified four common causes.

\textbf{Scattered code}.  To successfully edit code, developers may need information located in other files, blocking them from further progress editing. During the edit step, developers asked question such as ``How did this value got set?'' (D4), ``What is the difference between these two implementations?'' (D1), ``Where is this code located?'' (D5), and ``Does this function save the object property?'' (D7). To answer these questions, developers navigated and searched across multiple source files. The mechanics of searching and navigating between files created gaps within edit-run cycles. For example, D1 spent almost 20\% of their cycle time searching and navigating between files. This was to ``understand the lifetimes of session storage and make sure that [he] fully understood it'' before introducing changes to the session storage code.

\textbf{Unfamiliar third-party APIs}. Editing code using unfamiliar APIs sometimes interrupted the edit step as developers learned more by consulting internet resources. Developers asked questions such as ``What other APIs does this library have?'' (D1), ``What am I suppose to do to get the desire output?'' (D7), and ``Why does this API throw an error?'' (D8).  For example, D10 switched to reading documentation five times while working within an edit step. He gradually wrote code using the API, while spending 26\% of the cycle time consulting documentation.

\textbf{Disintegrated development environment}. Developers used a wide range of tools in their development environment, including a code editor, issue tracker, terminal, and note editor. These tools were often not integrated, causing gaps when developers switched to them. For example, three developers (D4, D5, D6) introduced a new third party library to the codebase during an edit step. Before using the libraries, they stopped editing, switched to the terminal, invoked the installation script for the library, and waited for the library to install before continuing to edit the source code. Between edit-run cycles, three developers (D4, D8, D11) periodically switched to the version control system to check code history and commit new changes.

\textbf{Waiting to compile}. Developers experienced gaps not only due to their immediate needs but also when waiting for operations to complete. For example, while waiting for their program to compile and the test cases to execute, three developers (D4, D5, D11) switched to check their issue tracker and other communication channels. This behavior did not increase the length of cycles by itself, as developers were already waiting for their program to run. However, it was a cause of gaps within run steps.

\begin{table*}[]
    \centering
    \caption{Design recommendations for enabling fluidity within edit-run cycles}
    \begin{tabular}{p{3.5cm}p{4.4cm}i{9cm}}
       \midrule
        Cause of Gap
        & 
        Design Recommendation
        & 
       \item[]Examples
        \\
         \toprule
        Scattered  relevant code 
         & 
         Reduce the overhead of searching and navigating multiple files. 
         &
            \item Help developers quickly locate and navigate between relevant code (e.g., REACHER \cite{latoza2011visualizing})
             \item Allow developers to create and navigate within task-specific views (e.g., Code Bubbles \cite{bragdon2010code}, Patchworks \cite{henley2014patchworks}).
         \\
         && \item[]\\
         Unfamiliar third-party APIs
         &
         Help developers quickly find documentation, code examples, and explanations of API errors. 
         &
         \item Offer in-situ auto complete and documentation for third party APIs (e.g, IntelliSense \cite{IntelliSense}).
         \item Support searching for and integrating code examples (e.g, Codeexchange \cite{martie2015codeexchange}, Strathcona \cite{holmes2005using}, and Codelets \cite{oney2012codelets}).
         \item Offer potential fixes and explanations of error messages (e.g, HelpMeOut \cite{hartmann2010}).
         \\
         && \item[]\\

         Disintegrated development environment 
         &
         Reduce the need to switch to different tools by seamlessly integrating them into the same environment.
         &
         \item Offer integrated support for committing and browse code changes  (e.g., Auto-git \cite{auto-git}). 
         \item Auto-detect and install third party libraries (e.g, Auto Install \cite{AutoInstall}).
         \item Offer integrated note-taking views for tracking task progress. 
         \\
         \bottomrule
    \end{tabular}
    \label{tab:consideration}
\end{table*}

\section{Limitations and threats to validity}

Our analysis of developers' edit-run workflow have several limitations and potential threats to validity. One potential threat to \textbf{external validity} is the representativeness of the developer activities that we analyzed. Our analysis of edit-run cycles drew from an existing dataset of professional developers at work~\cite{alaboudi2021exploratory}. All the videos in the dataset show professional developers with at least seven years of experience working on active open source projects. Developers worked within a diverse set of development environments, programming languages, and technology stacks. The work that developers did during the videos was eventually committed to the project.

A potential threat to \textbf{construct validity} is that the method we used to construct edit-run cycles did not actually capture edit-run cycles. To mitigate this threat, we first reviewed existing live-programming literature \cite{rein2018exploratory} and defined the activities constituting edit-run cycles. We then wrote a script that automated the construction of cycles based on these definition. Finally, we coded randomly selected cycles for all developers and checked that the constructed cycles started and ended in accordance to our definitions of edit-run cycles.  

Inherent to the nature of exploratory studies, our analysis was purely descriptive and does not establish cause and effect relationships between any of the measures we observed. Our work offers a characterization of the activities that occurred. Future work is required to examine the causality of relationships, particularly through controlled experiments. 

\section{Discussion}
Live programming environments aim to empower developers by creating a fluid edit-run experience where developers engage in short, frequent, and sequential edit-run cycles. We conducted the first empirical study investigating how developers' current edit-run behavior compares to the fluid ideal envisioned by live programming environments. We found that edit-run cycles were generally short when both debugging and programming and under 5 minutes in length.  Edit-run cycles in debugging were shorter, lasting on average one minute per edit-run cycle instead of three minutes for edit-run cycles in programming. We found that edit-run cycles in debugging were more frequent, where developers on average tried out and ran seven edits before fixing a defect. Edit-run cycles were less frequent in programming, with only two cycles per programming episode. This suggests that debugging work most closely embodies the rapid and fluid cycles envisioned by live programming. Moreover, as envisioned by live programming, developers spend most of their time in edit-run cycles editing, focusing primarily on editing a single file of code and testing the impact on the final output. Developers  who  used  a  traditional text  editor  had  69\% more edit-run cycles in each debugging episode, which were each  55\%  longer,  than  developers  who  used  an  IDE.

We found that gaps within and between edit-run cycles sometimes occurred, interrupting developers. Cycles with gaps were, on average, four times longer than those without. These gaps were caused mainly by developers engaged in satisfying information needs. A variety of studies of information needs examine in detail potential causes of these gaps, reporting questions that developers ask~\cite{Ko2007InformationNeeds, LaToza2010, Sillito2008}. While live programming systems traditionally focus on more directly supporting the edit or run step, our findings offer evidence of the importance of supporting developers' information seeking.
To better support fluidity in edit-run cycles, development environments must provide more rapid and integrated support for satisfying these needs. 

We propose three design recommendations for increasing the fluidity of developers' work within edit-run cycles (Table \ref{tab:consideration}). These recommendations broaden the traditional scope of the live programming environment, encompassing more of developers' work within development tasks. We also offer several examples of systems which embody techniques following these recommendations. 

One cause of gaps was searching across files for scattered code. Live programming environments should aim to reduce these gaps by creating better task-focused views, helping developers find and navigate between code more quickly. 

When working with unfamiliar third-party APIs, developers were distracted from their edit-run cycles as they searched for documentation, code examples, and explanations of error messages. Live programming environments might do more to integrate support for these needs within the programming environment, through in-situ documentation \cite{IntelliSense}, code examples \cite{martie2015codeexchange, holmes2005using, oney2012codelets}, and debugging support \cite{alaboudiHypotheses, hartmann2010}.  

Fluidity was also reduced in cases where developers switched between tools. Developers committed their progress to a version control system, installed libraries, and documented their progress and plans. Live programming environments might increase fluidity by integrating, or even automating, more of this functionality within the IDE.

%
\section*{Acknowledgments}
This research was funded in part by the National Science Foundation Grant CCF-1845508. 

\bibliographystyle{IEEEtran}
\bibliography{Debugging}

%

%

\end{document}